# Integrated Simulation Platform for Quantifying the Traffic-Induced Environmental and Health Impacts

**Xuanpeng Zhao, Corresponding Author**

Department of Electrical and Computer Engineering

University of California, Riverside, Riverside, CA 92521

xzhao094@ucr.edu

**Guoyuan Wu, Ph.D.**

Center for Environmental Research and Technology

University of California, Riverside, Riverside, CA 92507

gywu@cert.ucr.edu

**Akula Venkatram, Ph.D.**

Department of Mechanical Engineering

University of California, Riverside, CA 92521, USA

venky@engr.ucr.edu

**Ji Luo, Ph.D.**

Center for Environmental Research and Technology

University of California, Riverside, Riverside, CA 92507

ji.luo@ucr.edu

**Peng Hao, Ph.D.**

Center for Environmental Research and Technology

University of California, Riverside, Riverside, CA 92507

haop@cert.ucr.edu

**Kanok Boriboonsomsin, Ph.D.**

Center for Environmental Research and Technology

University of California, Riverside, Riverside, CA 92507

kanok@cert.ucr.edu

**Shaohua Hu,**

California Air Resources Board

1001 I St #2828, Sacramento, CA 95814

Shaohua.Hu@arb.ca.gov

**ABSTRACT**

Air quality and human exposure to mobile source pollutants have become major concerns in urban transportation. Existing studies mainly focus on mitigating traffic congestion and reducing carbon footprints, with limited understanding of traffic-related health impacts from the environmental justice perspective. To address this gap, we present an innovative integrated simulation platform that models traffic-related air quality and human exposure at the microscopic level. The platform consists of five modules: SUMO for traffic modeling, MOVES for emissions modeling, a 3D grid-based dispersion model, a Matlab-based concentration visualizer, and a human exposure model. Our case study on multi-modal mobility on-demand services demonstrates that a distributed pickup strategy can reduce human cancer risk associated with PM2.5 by 33.4% compared to centralized pickup. Our platform offers quantitative results of traffic-related air quality and health impacts, useful for evaluating environmental issues and improving transportation systems management and operations strategies.

**Keywords:** Simulation of Urban MObility (SUMO), Motor Vehicle Emissions Simulator (MOVES), environmental and health impact, curbside management

# INTRODUCTION

**Motivation**

In recent years, increased transportation-related activities have raised awareness and concerns about air pollution and adverse health effects. In 2022, the transportation sector was the largest producer of Greenhouse Gases (GHG) nationwide, accounting for approximately 28.2% of total U.S. emissions (U.S. EPA, 2022). To address the issues, a variety of emerging mobility technologies and services, such as connected and automated vehicles (CAVs), smart infrastructure and shared mobility, have been developed and deployed over the past decade (Sperling, 2018). For example, CAV technology has been widely studied to improve the sustainability of transportation systems, where a CAV can be driven by itself with the help of its on-board perception sensors, and also communicate with the other equipped vehicles (through vehicle-to-vehicle or V2V communications), roadside infrastructure (through vehicle-to-infrastructure or V2I communications), and the "Cloud" (Wang et al., 2020). Representative applications for urban scenarios are eco-approach and departure (Altan et al., 2017), (Katsaros et al., 2011). Besides the advanced technologies on the vehicle side, some researchers focus on the infrastructure side to improve the overall energy efficiency of the traffic system. Lee et al. proposed a cooperative vehicle intersection control system that enables cooperation between vehicles and infrastructure for effective intersection operation and management (Lee and Park, 2012), thus enhancing system throughput and environmental sustainability.

In addition to the advances in both the vehicle and infrastructure sides, emerging multi-modal mobility on-demand ($M^3OD$) services such as micro-mobility and ride-hailing have not only unlocked novel opportunities for urban transportation but also introduced new challenges to users, service providers, and public transportation agencies alike. For example, increased curbside activities due to the prevalence of multi-modal Mobility as a Service (MaaS) have not only created congestion for traffic of different modes along the curbs or on the sidewalks, but also formed potential bottlenecks that may affect upstream on-road vehicular flows. From an environmental perspective, traffic congestion near the curbside would lead to energy waste and excessive tailpipe emissions, thus forming hotspot(s) with high pollutant concentration. Even worse, due to the high-volume pedestrian and/or other non-motorized traffic on the sidewalk, many more safety risks will be raised, and detrimental health impacts would be imposed on those vulnerable road users. However, most research related to these emerging $M^3OD$ services has been focused on accessibility, safety, and congestion impacts, but much less attention has been raised from the perspective of their resultant environmental and health impacts. In particular, how well-designed curbside management strategies would affect roadway (vehicular) traffic as well as sidewalk (pedestrian or other micro-mobility) traffic is still an open research question. In addition, how these strategies will influence air quality and vulnerable road users' exposure to motor vehicular pollutants and toxins is a critical concern for local governments (e.g., cities, MPOs), especially for disadvantaged communities.

To address the aforementioned gaps, we develop an integrated simulation platform in this study (as shown in Figure 1) which is able to: 1) model a multi-modal transportation system with high resolution, including roadway network, motorized vehicles (such as passenger cars, trucks, and shuttles), and non-motorized transportation modes (such as pedestrians, bicyclists, or even other micro-mobility travelers); 2) model traffic related pollutant emissions which include tailpipe emissions and even brake/tire worn emissions; 3) model the air quality impact of these emissions air quality dispersion; 4) model the potential exposure to those vulnerable road users; 5) implement different curbside management strategies in response to multi-modal mobility on-demand ($M^3OD$) services; and 6) evaluate the system performance especially in terms of environmental footprints and health impacts.

**Organization of the Paper**

The rest of this paper is organized as follows: the next section reviews the existing literature related to microscopic traffic simulators, vehicular emissions models, tailpipe pollutant dispersion models and their applications to urban traffic scenarios. Major efforts in developing the integrated modeling platform are presented in Section 3, where each module is elaborated. Based on the real-world network in the City of Riverside, Section 4 describes a case study on M$^3$OD services and provides quantitative results of tailpipe emissions and human exposure for different curbside management strategies. The last section concludes this study with potential improvements for future research.

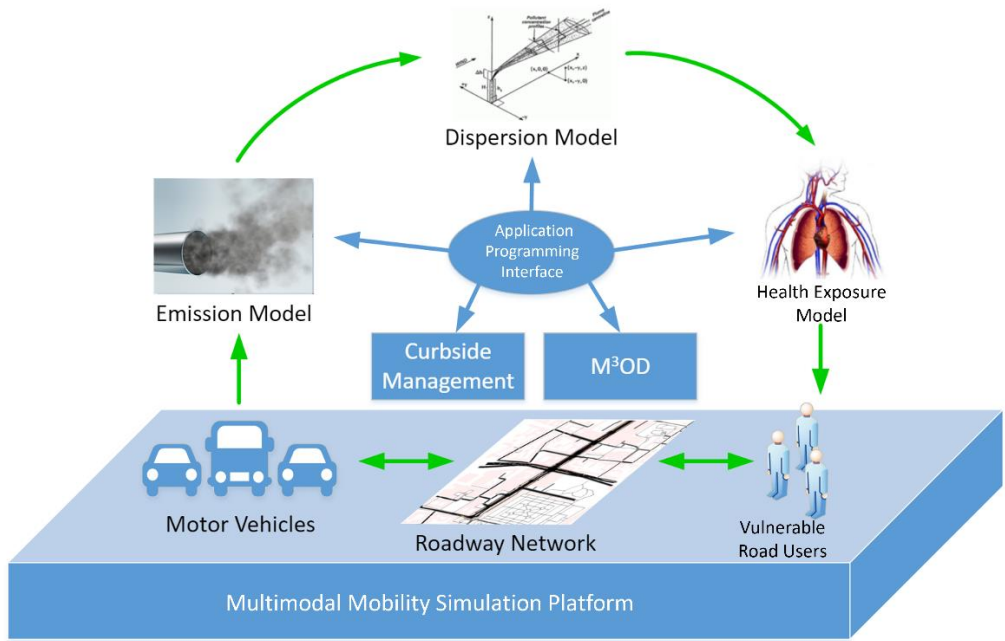

**Figure 1. The concept of integrated modeling for quantifying the environmental and health impacts related to M$^3$OD transportation services.**

## BACKGROUND AND RELATED WORK

### Microscopic Traffic Simulator

Currently, there are several microscopic simulators available to support modeling traffic scenarios in a realistic setting, including the roadway network, motor vehicles and non-motorized road users (e.g., pedestrians, bicyclists). Among them, PTV VISSIM (Fellendorf and Vortisch, 2010) and Aimsun (Aimsun, 2022) are two major commercial simulation tools. Specifically, VISSIM is a behavior-based multi-purpose microscopic simulation that can be linked with MATLAB through the Component Object Model (COM) interface or with C/C++ via dedicated application programming interfaces (APIs). Aimsun is a hybrid traffic modelling simulator which allows simultaneous application of multi-model analysis with large networks. On the other hand, SUMO (Behrisch et al., 2011) is an open-source traffic simulator that has been used for a variety of applications, such as dynamic navigation, traffic surveillance systems evaluation, and traffic light algorithm development (Krajzewicz et al., 2012). In addition, SUMO provides an API, Traffic Control Interface (TraCI), to facilitate the interaction with external applications through a socket (bidirectional) connection. It should be noted that SUMO also includes a few emissions models, e.g., the Handbook Emission Factors for Road Transport (HBEFA), PHEM (Passenger Car and Heavy-Duty Emission Model), developed for European vehicle fleets.

### Microscopic Emission Model

Vehicular emissions models estimate emissions rates and/or emission factors of motorized vehicles based on different traffic conditions and driving cycles. They can be classified into three main types, i.e., microscopic models, macroscopic models, and average velocity-based statistical models. In this study, we focus on the microscopic models derived from the relationship between the second-by-second vehicle trajectories and emission rates. The Comprehensive Modal Emissions Model or CMEM is able to predict second-by-second tailpipe emissions and fuel consumption based on different modal operations from in-use vehicle fleets (Barth et al., 2000). The calculation method fully considers the power and speed of the engine, to accurately reflect the emission characteristics of the vehicle, which belongs to transient physical models. Another well accepted model is the Virginia Tech microscopic (VT-Micro) model that was developed using chassis dynamometer data on light duty vehicles and trucks (Rakha et al., 2004). A polynomial regression model on key vehicle dynamics, e.g., speed, acceleration, and power, is set up to estimate non-steady-state emissions. As aforementioned, HBEFA is adopted in SUMO and it is widely used for fleets in European countries (INFRAS, 2009). Based on traffic activities, HBEFA can provide emission factors by a) type of emission (e.g., hot run, cold start), b) vehicle category (e.g., passenger cars, heavy duty vehicles, buses), c) year (1990-2050 for most countries), d) pollutants including CO, HC, NOx, PM, $CO_2$, $NH_3$ and $N_2O$.

Over the past decade, U.S. Environmental Protection Agency (EPA) has been developing a state-of-the-art vehicular emissions model, called Motor Vehicle Emissions Simulator (MOVES) (U.S. EPA, 2009). By applying the binning strategy, MOVES aims at estimating vehicular emissions at multiple scales, i.e., microscopic (for individual vehicles), mesoscopic (based on link-level traffic data), and macroscopic (i.e., aggregated inventory for a region or even the entire nation). The open database and model structure of MOVES increase its transferability, allowing other stakeholders to collect their own datasets that represent local traffic conditions, vehicle mix and driving trajectories to estimate the specific tailpipe emissions inventory. In particular, the open database in MOVES stores the base emissions rates of different criteria pollutants for different vehicle types, vehicle ages, and operating mode bins (depending on speed, acceleration, and vehicle specific power).

**Dispersion Models and Their Application to Urban Scenarios**

Several dispersion models have been developed to examine the impact of vehicle emissions on urban air quality. Lefebvre et al. presented an integrated model framework consisting of a measurement interpolation model, a bi-Gaussian plume model, and a canyon model to simulate urban traffic scenarios at the street level (Lefebvre et al., 2013). Shi et al. leveraged a CFD-based model to simulate scenarios with street canyons and visualize exhaust emissions of moving vehicles using dynamic mesh updating which could even capture the vehicle movement-induced turbulence effects (Shi et al., 2020). Damoiseaux and Schutter developed a Line Source Gaussian Puff (LSGP) model capable of estimating distributions of gaseous pollutants in the vicinity of a freeway and applied it to assessing real-time traffic control, e.g., variable speed limit (Damoiseaux and De Schutter, 2021). Nevertheless, LSGP is not suitable for microscopic (at the individual vehicle level) simulation, as it is a line source model. Zegeye proposed a 2D point-source dispersion model, which may update the pollutant concentration on a grid basis (Zegeye, 2011). This model requires minimal computational resources, which makes it suitable for online estimation of environmental impacts. However, the lack of support in theory dims the validity of results from Zegeye's model, and it is questionable to use a simplified decay function for vertical dispersion simulation without careful consideration of the mass conservation.

In principle, any one of these dispersion models can be incorporated into the simulation platform described in this paper. Instead, we have incorporated a 3-D grid model, which while not including the details of dispersion in an urban area, incorporates the primary processes that allow us to evaluate the impact of alternative transportation strategies. The model captures the unsteady aspects of emissions from moving vehicles and their subsequent dispersion in a framework that facilitates computational efficiency, a requirement for an integrated platform. While the model, in its current form, does not include the effects of urban street canyons, it can provide a time series of concentrations at specified receptors, and concentration gradients associated with moving vehicles.

**INTEGRATED MODELING PLATFORM**

In this study, we leverage the capacity of SUMO and build up an integrated modeling platform (as depicted in Figure 1) by adding other key modules, including MOVES, a 3-dimensional (3D) an unsteady grid-based dispersion model, a human exposure model, a Matlab-based concentration visualizer, and heuristic strategies for $M^3OD$ services (via TraCI).

**Overall Workflow**

Figure 2 presents the workflow developed for the modeling platform to quantify the environmental and health impacts related to $M^3OD$ services. As illustrated in the figure, we firstly create a multi-modal traffic network in SUMO, including taxis or Uber/Lyft vehicles, background traffic (i.e., passenger vehicles) and pedestrians waiting for services. Then, the second-by-second vehicular tailpipe emissions are estimated by the coded MOVES model. With the developed dispersion module, grid-wide pollutant concentrations are estimated based on both emissions and meteorological conditions. And the results can be visualized online through a custom-built visualization tool using Matlab. The instant and accumulative exposure to specific pollutant (e.g., NOx, PM) for each individual vulnerable road user (VRU) can be estimated, depending on his/her instant location (i.e., in which grid at each time step) and trajectory, as well as characteristics related to gender and age (such as height, breathing rate).

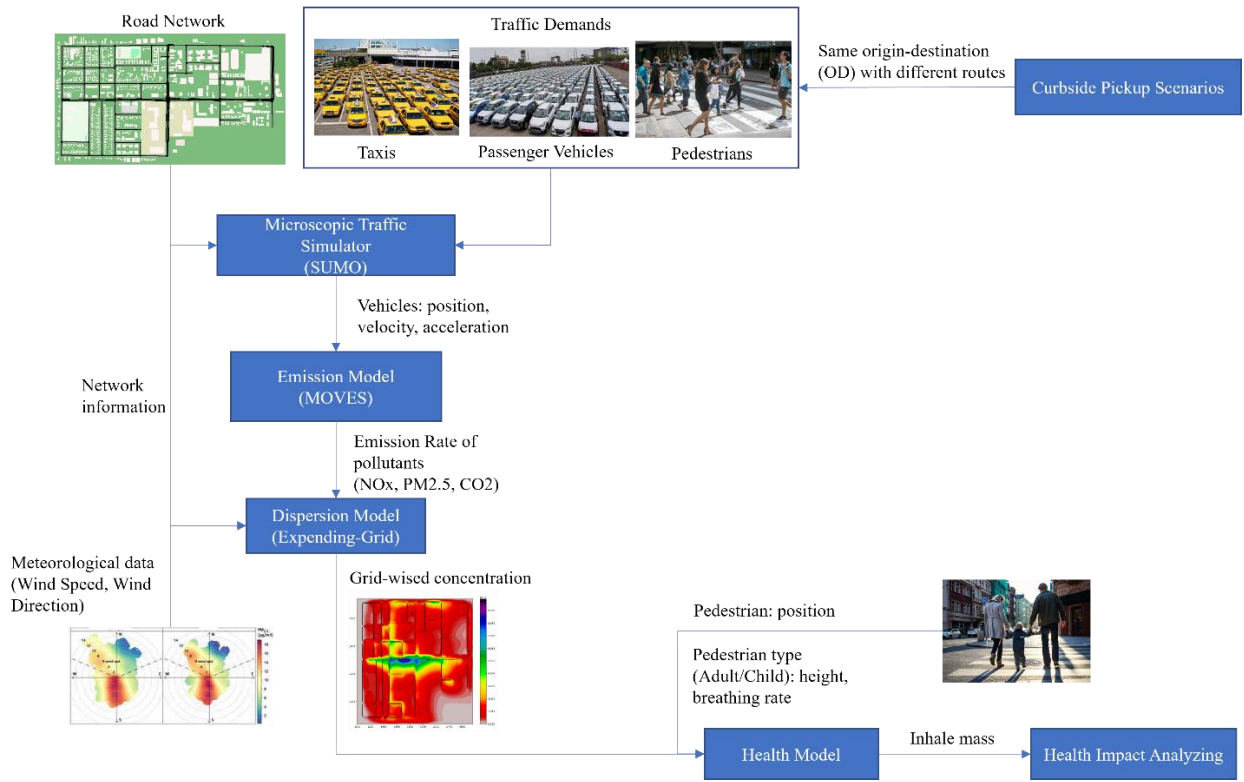

**Figure 2. The workflow to run the integrated modeling platform.**

**Platform Inputs and Outputs**

This section describes the platform input variables, and the outputs provided by the integrated modeling platform.

**1. Input**

*Geographic area roadway network*

SUMO provides a python script called OsmWebWizard.py which can select the real-world region from the Open Street Map (OSM) to generate the target simulation roadway network without laborious efforts. Another script called netconvert.py provided by SUMO can convert OSM to SUMO compatible networks.

Besides, users can also use a network editor named "Netedit" included in SUMO to create and modify self-defined network input compatible with SUMO.

*Traffic activity of vehicles and pedestrians*

SUMO provides a python script called RandomTrips.py which can generate a set of random trips in a given network that would apply to both vehicles and pedestrians. Self-defined vehicle profiles and pedestrian profiles are also allowed.

In general, vehicle profiles have a list of vehicles, and each vehicle has basic attributes including departure time and position, arrival position, predefined route, acceleration, minimum gap, and vehicle length, etc. Specifically, special vehicle types can be given special attributes. For example, drop-off and pick-up duration can be defined for a taxi type vehicle.

Besides the basic attributes are similar to vehicle attributes, pedestrian attributes divide predefined routes into routes of walking and routes of taking taxis by the location waiting for taxis.

In particular, to calculate inhalation, we added a "type" attribute to the pedestrian profile to indicate whether the pedestrian is an adult or a child.

*Meteorological condition*

Meteorological condition is composed by wind speed and wind direction of each cell in the map.

**2. Output**

*Vehicle information*

The vehicle information log is a JSON file, including type, location, speed, acceleration, heading, emissions rate and energy consumption rate of each individual vehicle at each time step. With post-processing, we can assess vehicular performance in terms of safety, mobility, and environmental sustainability.

*Traffic-related concentration*

The concentration log is a file with the extension ".npy" that stores concentration matrices for each pollutant of interest at every time step. Each matrix has dimensions of height, width, and length and can be further analyzed or replayed as needed.

*Pedestrian exposure*

This includes demographic characteristics (e.g., child/adult), location, speed, exposure of each individual pedestrian at each time step. The output file can be used to evaluate the quality of $M^3OD$ service (e.g., waiting time) and assess human exposure to the pollutants of interest either individually or in an aggregated manner.

**Platform Key Modules**

1. **Key Module 1 – SUMO**

SUMO mainly serves as a microscopic traffic simulator handling vehicles and pedestrian moving behavior using default control algorithms. Besides this, SUMO also provides powerful APIs to extend its capability. For those interested in further exploring the functionalities in SUMO, please refer to https://www.eclipse.org/sumo/.

2. **Key Module 2 – MOVES**

The original MOVES model developed by the U.S. EPA is very comprehensive and not suitable for on-line interaction with microscopic traffic simulation. In this study, we develop an alternative approach to simplify the application of MOVES to simulation while keeping reasonable fidelity similar to the original MOVES model. Figure 3 depicts the workflow of MOVES plug-in development for SUMO. Similar procedures can be applied to the development of other microscopic simulation tools (e.g., VISSIM). There are two major procedures (starting from the upper left corner): a) acquiring emission rate tables from MOVES; and b) calculating operating mode (OpMode) for each vehicle at each time step in the simulation.

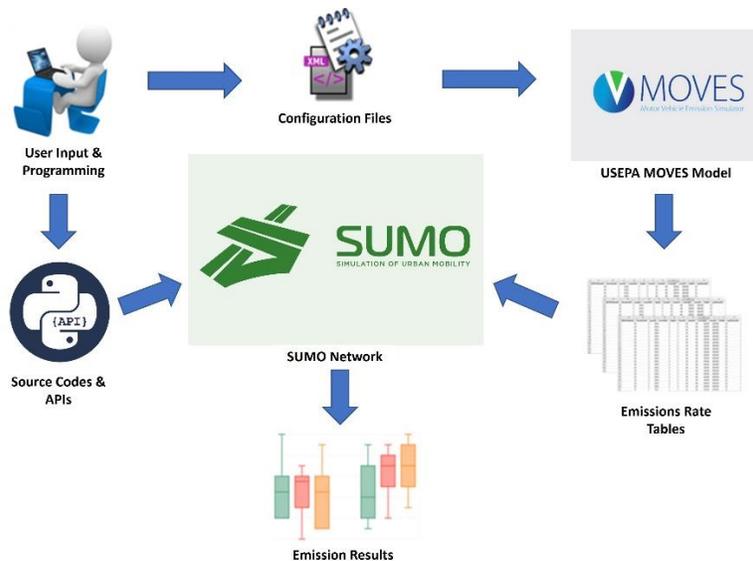

**Figure 3. The workflow for developing the MOVES plug-in in SUMO.**

*Acquiring Emission Rate Tables*

To retrieve the customized emission rate tables (from MOVES), we firstly input the network model information to the MOVES model, such as geographic region (e.g., Riverside in California), calendar month and year to be modeled (e.g., June 2022). In the meantime, we prepare a set of configuration files that can be linked to the MOVES database, including vehicle population/activity, fuel type/engine technology, vehicle inspection/maintenance program and meteorological statistics. Once all the input data files are ready, MOVES can be executed and output emission rate tables for different source types (e.g., passenger car, truck), considering various factors, such as vehicle model year distribution, fuel type/engine technology market share, and temperature and/or humidity adjustment.

*Calculating OpMode Distributions*

In SUMO, TraCI can access second-by-second vehicle trajectories (including both speeds and accelerations) and road grade (if any). With this activity data for each vehicle and roadway geometry, as well as the information on vehicle class and weight, the vehicle specific power (VSP) characteristics (in kWatt/tonne) can be calculated by (U.S. EPA, 2010):

$$VSP = \left(\frac{A}{M}\right) \cdot v + \left(\frac{B}{M}\right) \cdot v^2 + \left(\frac{C}{M}\right) \cdot v^3 + (a + g \cdot sin\theta) \cdot v, \tag{1}$$

where $A$, $B$ and $C$ are the road-load related coefficients for rolling resistance ($kW \cdot sec/m$), rotating resistance ($kW \cdot sec^2/m^2$) and aerodynamic drag ($kW \cdot sec^3/m^3$), respectively; $v$ is the vehicle speed ($m/sec$); $M$ is the mass of vehicle (metric ton); $g$ is the acceleration due to gravity (9.8 $m/sec^2$); $a$ is the vehicle acceleration ($m/sec^2$); and $\theta$ is the (fractional) road grade. Default values of these parameters are provided in (U.S. EPA, 2010). After the VSP values are calculated, they will be binned according to the MOVES' vehicle operating mode (OpMode) bin definition given in (Wang et al., 2021). With the emission rate tables coded in SUMO, the energy consumption and pollutant emissions can be estimated in either disaggregate (e.g., second-by-second for each vehicle) or aggregate in the spatial-temporal manner.

**Operating Modes for Running Exhaust Emissions**

| VSP Class (kW/tonne) | Speed Class (mph) 1-25 | 25-50 | 50+ |
|---|---|---|---|
| 30+ | 16 | 30 | 40 |
| 27-30 | | | |
| 24-27 | | 29 | 39 |
| 21-24 | | 28 | 38 |
| 18-21 | | | |
| 15-18 | | | 37 |
| 12-15 | | 27 | |
| 9-12 | 15 | 25 | |
| 6-9 | 14 | 24 | 35 |
| 3-6 | 13 | 23 | |
| 0-3 | 12 | 22 | 33 |
| < 0 | 11 | 21 | |

21 modes representing "cruise & acceleration" (VSP>0)

PLUS
2 modes representing "coasting" (VSP<=0)

PLUS
One mode each for idle, and decel/braking

Gives a total of 23 opModes

**Figure 4. The operating mode (OpMode) binning scheme defined in MOVES.**

## 3. Key Module 3 – Dispersion model

The concentrations resulting from the emissions from vehicles moving in the domain of interest are computed with an unsteady 3D grid-based model based on the mass conservation equation.

$$\frac{\partial q}{\partial t} + u(t)\frac{\partial q}{\partial x} + v(t)\frac{\partial q}{\partial y} = \frac{\partial}{\partial z}\left(K(z)\frac{\partial q}{\partial z}\right) + \frac{\partial}{\partial x}\left(K_x\frac{\partial q}{\partial x}\right) + \frac{\partial}{\partial y}\left(K_y\frac{\partial q}{\partial y}\right) + \frac{S(x,y,z,t)}{\rho_a}, \quad (2)$$

where $q$ is the mixing ratio of the pollutant; and $u(t)$ and $v(t)$ are the horizontal velocities along the orthogonal $x$ and $y$ coordinates used in the numerical solution of the equation. These velocities are taken to be constant over the domain but vary hourly. $K(z)$ is the eddy diffusivity in the vertical direction, which is modeled using Monin-Obukhov similarity theory (Businger et al., 1971), and $K_x$ and $K_y$ are horizontal diffusivities assigned nominal values.

The emission rate, $S(x, y, z, t)$, corresponds to the emissions from vehicles in the grid squares laid over the domain during $\Delta t$, and $\rho_a$ is the density of air. The emissions are updated every time step to account for the movement of vehicles.

Equation (2) is solved using the method of fractional steps, in which each component of the equation is updated over a time step, $\Delta t$, using the sequence 1) emission, 2) advection or horizontal transport by the wind, 3) Vertical diffusion, 4) horizontal diffusion in the x-direction, followed by 5) horizontal diffusion in the y-direction.

Advection is solved using a semi-Lagrangian technique in which the mixing ratio is updated by computing a backward trajectory from the center of the grid square over $\Delta t$, and then computing $q$ at the location of the starting point of the trajectory using bi-linear interpolation. So, the advection component

$$\frac{\partial q}{\partial t} + u(t)\frac{\partial q}{\partial x} + v(t)\frac{\partial q}{\partial y} = 0, \quad (3)$$

is solved using the equivalent equation,

$$q(x, y, z, t + \Delta t) = q(x - u\Delta t, y - v\Delta t, z, t) \tag{4}$$

The $q$ at $(x - u\Delta t, y - v\Delta t)$ are obtained through bi-linear interpolation of the $q$ field at time, $t$.

The vertical and horizontal diffusion terms are solved separately using a finite difference equation that is implicit in time. This finite difference representation results in a tridiagonal system of linear equations that are readily solved using explicit methods.

The meteorological inputs consist of the wind field, $U = (u, v)$, the wind direction $\theta_w$, the friction velocity $u_*$, the Monin-Obukhov length $L$, the mixed layer height $z_i$, the roughness length $z_0$, and the measurement height $z_{ref}$. The micrometeorological variables, which are derived from AERMET, the meteorological processor of AERMOD (Cimorelli et al., 2005), are used to compute the vertical eddy diffusivity, $K(z)$ in Equation (2). The eddy diffusivities can be enhanced to account for vehicle-induced turbulence. We can also account for street-canyon enhancement of concentrations using the method described in (Schulte et al., 2015). These features have not yet been incorporated in the current version of the dispersion model.

## 4. Key Module 4 – Matlab-based Online Visualizer

SUMO has its own graphical user interface (GUI) showing the road network, real-time traffic and pedestrian flows, but not the emissions or concentrations. In this project, we develop a Matlab-based routine for visualizing the real-time concentration of the area of interest using colormap, along with the simulation run in SUMO. To achieve this, we first read and display the road network from the net file of SUMO. Then, we create grids based on the size of the entire network and the defined grid size. The side length of the entire mesh area is extended to be 10% larger than the target road network to avoid any margin problem. Finally, we display the concentration matrix by "nipy_spectral" colormap. There is a color bar which indicates the mapping of data values, and its range keeps updating based on the latest concentration matrix. Figure 5 presents an example of a concentration map (overlaying the roadway network) output from the Matlab-based Visualizer.

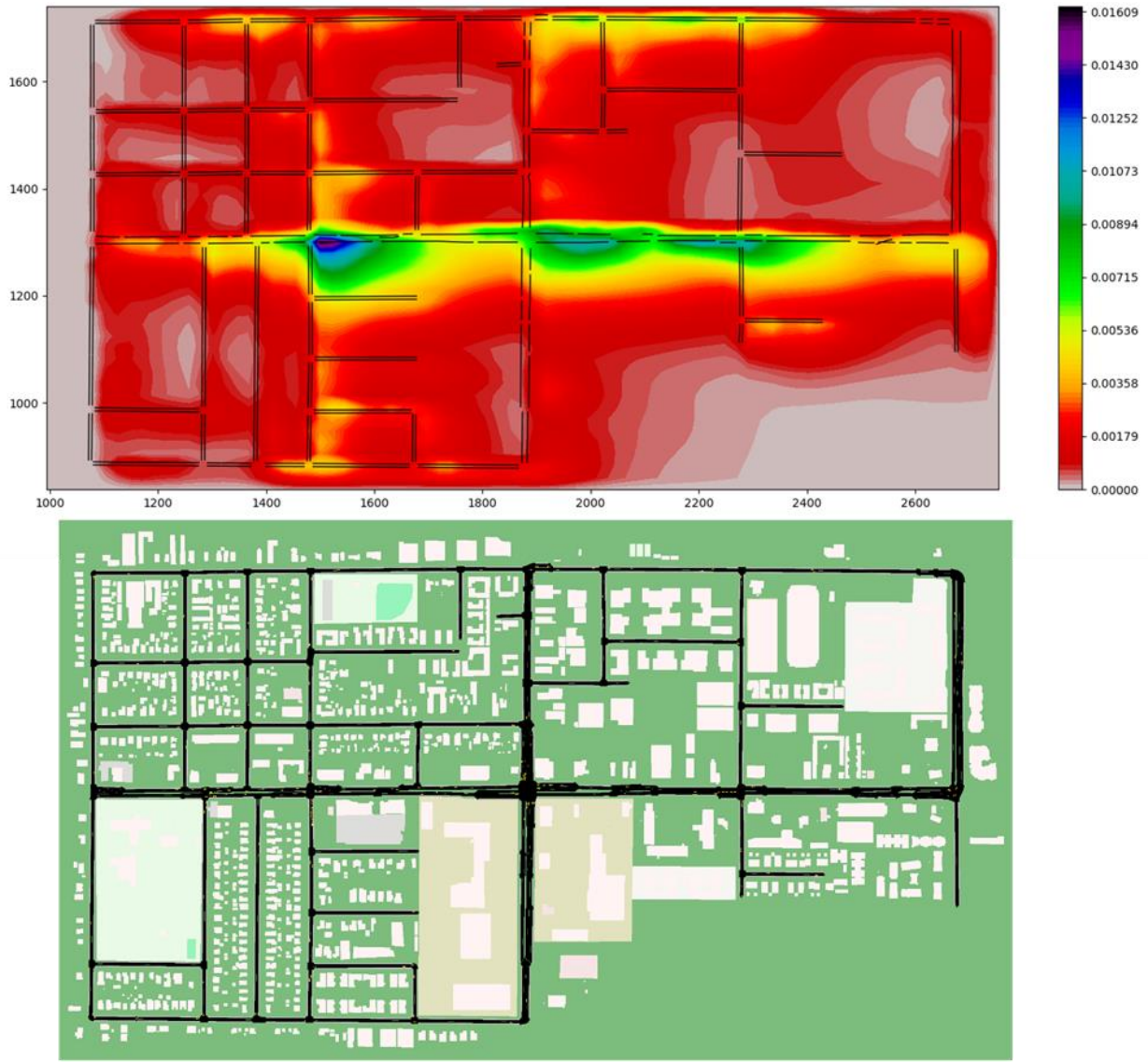

**Figure 5. A screenshot of the Matlab-based Visualizer for an example simulation run in SUMO.**

**Key Module 5 – Human Exposure Model**

In this research, human exposure refers to the amount of pollutant inhaled by a person subject. To assess the pollutant exposure, *inhaled mass* (IM) is used as the metric and is calculated by the following Equation (Bennett et al., 2002). Assuming a pedestrian subject $i$ is located within grid$(x_{jc}, y_{jc})$ at time step $k$, then

$$IM_i(k) = conc(x, y, k) \cdot \Delta t \cdot BR_i(k), \tag{5}$$

where $conc(x, y, k)$ is the pollutant concentration ($\mu g/m3$) in grid$(x, y)$ at time step $k$; $\Delta t$ is the time step; and $BR_i(k)$ denotes the breathing rate (assume to be constant in this study) of the $i$-th subject exposed to the pollutant at time step $k$. Breathing rates of different age groups can be referred to *U.S. EPA Exposure Factors Handbook* (U.S. EPA, 2011). In this study, we assume a population-wide average adult breathing rate to be 17 $m^3/day$. Also, we concern more on human exposure to PM$_{2.5}$, as they are associated with a range of health risks for many population groups (Weichenthal et al., 2012).

In the cancer risk assessment, we focus on the risks associated with traffic pollutant PM$_{2.5}$. The cancer risk equations presented below are based on the *Hotspots Analysis and Reporting Program (HARP) Air Dispersion Modeling and Risk Tool* (California Air Resources Board, 2022). All of the parameters used in these equations are chosen based on the acceptable ranges or adapted from existing research, to ensure valid assessment:

$$C_{air} = \frac{IM_{avg}}{BR \times t} \tag{6}$$

$$Dose = C_{air} \times \frac{BR}{BW} \times AF \times EF \times 10^{-6} \tag{7}$$

$$Cancer\ Risk = Dose \times PF \times \frac{ED}{AL} \times ASF \times CF, \tag{8}$$

where $Cancer\ Risk$ is defined as the risk a hypothetical individual faces of developing cancer, if exposed to carcinogenic emissions from a particular source for a specific duration; this risk is defined as an excess risk because it is above and beyond the background cancer risk to the population; and the cancer risk is expressed in terms of risk per million exposed individuals. In Equation (6), $C_{air}$ is the concentration of the target pollutant (in the unit of $\mu g/m^3$). $IM_{avg}$ is the average inhaled mass calculated from the dispersion model in $\mu g$. $t$ represents the amount of time that an individual is exposed to a toxic pollutant during the commute. In Equation (7), $Dose$ is the daily amount of a toxic pollutant (in the unit of $mg/kg/day$) that the human body absorbs. $BR$ and $BW$ refer to the breathing rate and body weight, respectively. The breathing rate to body mass ratio BR/BW is set to be 233 ($L/kg/day$) for adults. $AF$ is the absorption factor which is defined as 1 for adults. $EF$ is the exposure frequency which is set 250/365 for a working period of 250 days per year. In Equation (8), $PF$ represents potency factor for diesel particulate matter and is defined as 1.1 $(mg/kg/day)^{-1}$, $PF$ for diesel particulate matter is applied here as a conservative estimate due to lack of potency data for gasoline exhaust; $ED$ and $AL$ measure exposure duration and average lifespan, respectively. The ED/AL ratio is set 20/70 for a 20-year work period and 70-year lifespan. $ASF$ stands for age sensitivity factors and is defined as 1 for adults. CF is the hourly fraction spent on commuting during a day represented as a percentage (i.e., $t/24 \times 100\%$).

**CASE STUDY**

To showcase the capability of the integrated modeling platform, we consider a M³OD service scenario within an urban district and deploy a grid-based dispersion model to evaluate the health impact on customers under two different curbside pickup strategies: centralized and distributed.

**Simulation Scenarios**

We use osmWebWizard.py provided by SUMO to extract the roadway network of a target area in Riverside, CA, as shown in Figure 6. This roadway network is bounded by W Linden Street (Northmost), 12th Street (Southernmost), Kansas Avenue (Westernmost), and Iowa Avenue (Easternmost). To facilitate pedestrian demand generation, we further add the sidewalk for each edge in the network. To enable the dispersion modeling, we divide the network into a 353×185 grid matrix with the grid size of 5×5m. As for the height (along the vertical axis), we define six layers, i.e., 0.3, 1.2, 4.8, 19.2, 76.8, and 200m. The selected height values follow a geometric progression where each is four times greater than the previous one except for the last value (i.e., 200m) which is considered as the boundary layer. This progression ensures a well-distributed sampling of pollutant concentrations at different altitudes, efficiently capturing the dynamics of pollutant dispersion in the vertical direction. In this case, at every time step the platform can generate 353×185×6 matrices for the concentrations of each emitted pollutant. The entire simulation duration is set at 3600 seconds with a time step of 1 second. In addition, the dispersion model parameters are established as follows: Monin-Obukhov length *L* of -200, the mixed layer height $z_i$ of 200m, the roughness length $z_0$ of 0.1m, and the measurement height $z_{ref}$ of 5m.

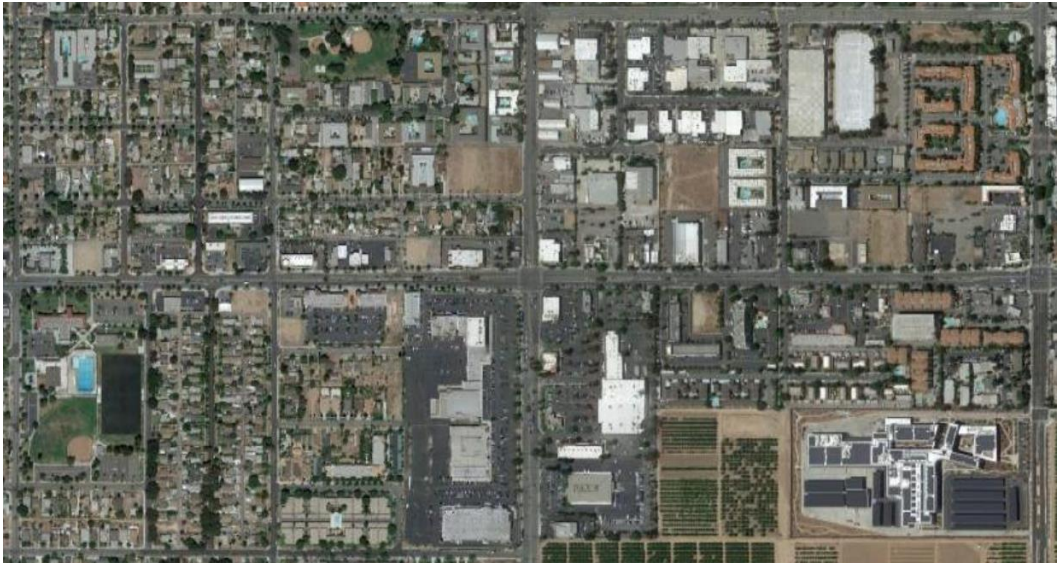

**(a) View of study network in Google Map**

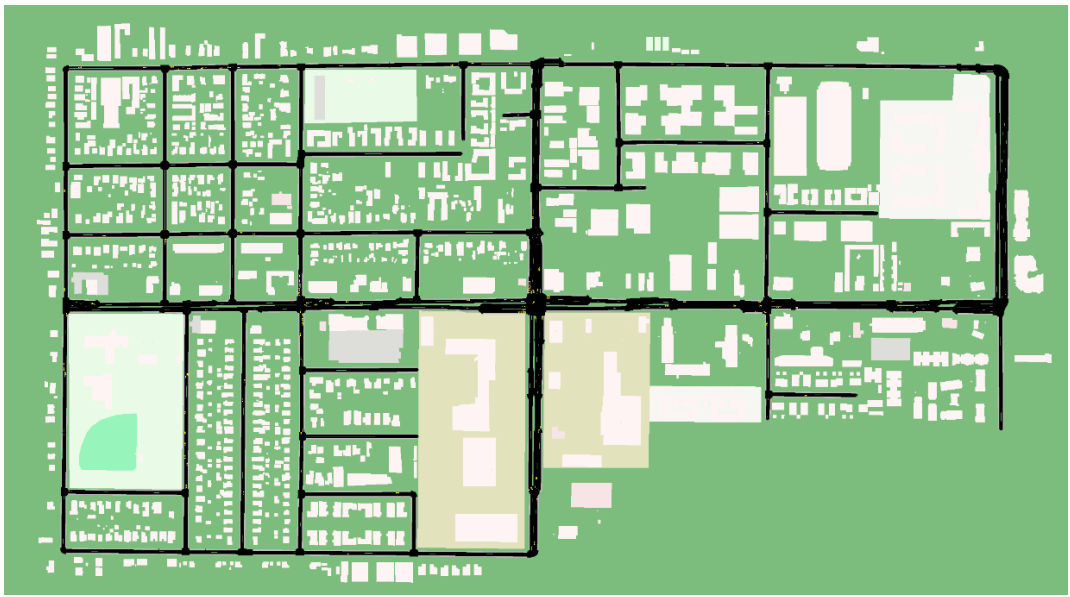

**(b) View of study network in SUMO**

**Figure 6. Illustration of the study network.**

To efficiently obtain reliable simulation results, we consider the balance between spatial resolution and computational efficiency when selecting the grid size. We evaluate the average computational time over a range of grid sizes, including 1x1, 3x3, 5x5, 10x10, 12x12, 15x15, and 22x22 as shown in Figure 7. Based on this analysis, a 5x5 grid size is selected in this study as it demonstrates a good balance (i.e., "elbow effect") between spatial resolution and computational efficiency.

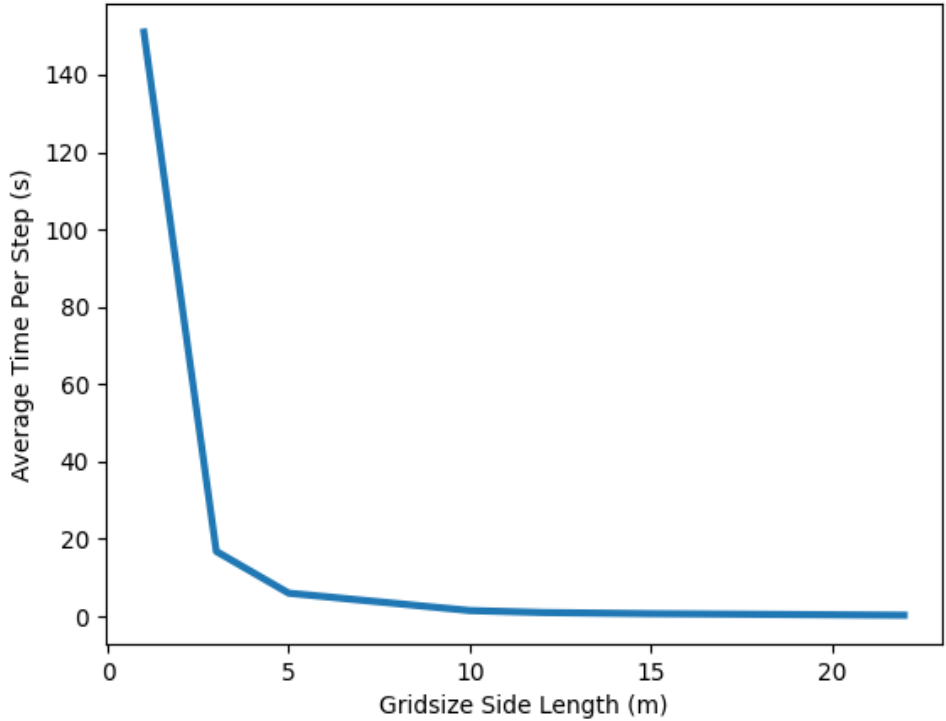

**Figure 7. Computational time per time step for various grid sizes.**

In the simulation, we set up the vehicle mix to be 95% of passenger vehicles and 5% of taxis/TNC vehicles. The passenger vehicles are considered background traffic, while taxis/TNC vehicles can receive pick-up requests, pick up passengers at their waiting locations and drop off them at their destinations. All the vehicles and pedestrians are randomly generated. Unlike the passenger vehicles which are spawned and terminated once completing the predefined routes, SUMO provides taxis/TNC vehicles with the idling algorithm option to enable them to continue driving randomly until the next request is received. In addition, taxis/TNC vehicles can specify the drop-off durations and pick-up duration, as shown in Table 1.

**Table 1. Parameter Settings for Simulation Scenarios in SUMO**

| Vehicle Type | Taxi | Passenger Vehicle |
|---|---|---|
| Number | 120 | 2280 |
| Minimum Gap (m) | 5 | 5 |
| Maximum Acceleration (m/s2) | 4 | 4 |
| Maximum Deceleration (m/s2) | -5 | -5 |
| Length (m) | 5 | 5 |
| Pick up Duration (s) | 5 | -- |
| Drop off Duration (s) | 10 | -- |

We assume the traffic volume per hour in this urban district is 2400 vehicles per hour (VPH). Based on the predefined vehicle mix, the number of passenger vehicles is 2280 and the number of taxis/TNC vehicles is 120. Pedestrians also have random origins and destinations, similar to the configuration of vehicles in SUMO, but they need to specify the locations where they should wait for pickups. Moreover, pedestrians can only access sidewalks and crossings, rather than motor roadways. It is noted that pedestrian characteristics, such as height and breathing rate, can be customized to model different groups, e.g., children and adults. In this study, we only randomly generate 240 adults.

To evaluate the human health impacts caused by different pickup strategies, we set up two scenarios: centralized pickup and distributed pickup. As for the centralized pickup, we create a central station at one road segment, as shown in Figure 8(a), highlighted with a red box. In this scenario, all pedestrians have to walk to the station and wait for pickups as shown in Figure 8(b) as blue dots, and all for-service vehicles have to meet respective passengers at the same station. In the second scenario, besides the station defined in Scenario 1, we create another 7 stations round as shown in Figure 8(c). Pedestrians are randomly distributed to these 8 stations to wait for pickups. The wind speed is set to be 10m/s blowing from northwest to southeast, which is a typical meteorological condition in Riverside, CA in Summer.

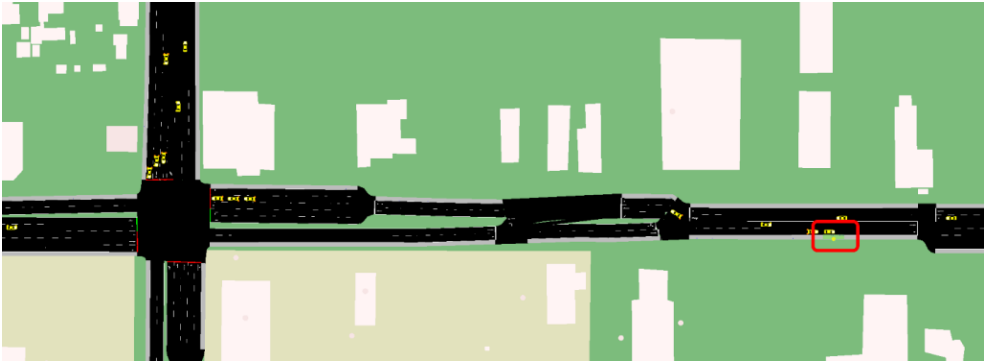

**(a) Centralized pickup**

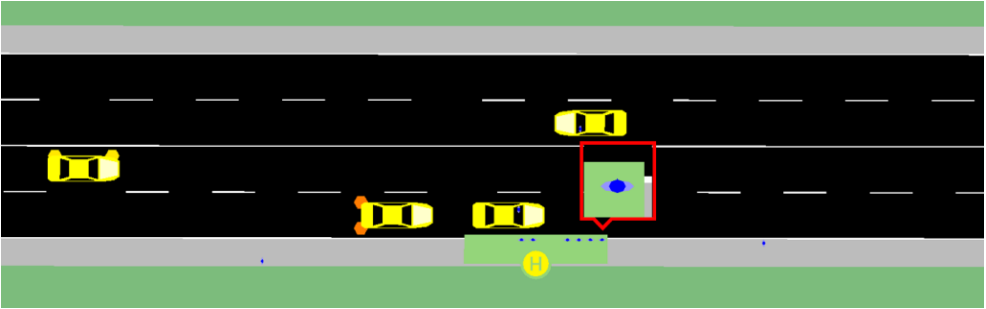

**(b) Zoomed in station in centralized pickup scenario**

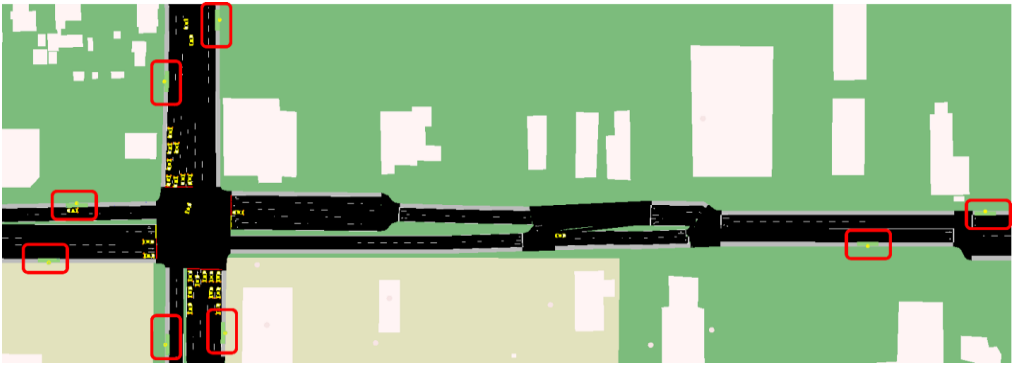

**(c) Distributed pickup**

**Figure 8. Station configuration for different curbside pickup strategies.**

## Results

### 1. Vehicular Emissions

Table 2 compares the emissions results between two pickup strategies. As can be observed from the table, the total emissions of NOx and PM$_{2.5}$ for the distributed pickup scenario are slightly higher than those of the centralized pickup scenario. This is mainly due to the increase in vehicle-mile-traveled (VMT). After normalization by the distance, there is no significant difference in total emission factor (i.e., Total Emissions/VMT) between two strategies, ranging from -0.7% to -0.9%.

**Table 2. Vehicular Emissions Comparison for Different Pickup Strategies**

| Scenario | | Centralized | Distributed | Difference |
|---|---|---|---|---|
| Vehicle-mile-traveled (mile) | | 2801 | 2879 | 2.8% |
| Average Velocity of Background Passenger Vehicles (mile/hour) | | 23.01 | 22.85 | -0.7% |
| Average Velocity of Taxis (mile/hour) | | 18.14 | 19.29 | 6.34% |
| Total Emission of Passenger Vehicles (g) | NOx | 243.54 | 244.16 | 0.3% |
| Total Emission of Taxis (g) | | 138.60 | 145.33 | 0.5% |
| Total Emission (g) | | 382.15 | 389.48 | 1.9% |
| Total Emission/VMT (g/mile) | | 0.13642 | 0.1352 | -0.9% |
| Total Emission of Passenger Vehicles (g) | PM$_{2.5}$ | 2.20 | 2.20 | 0% |
| Total Emission of Taxis (g) | | 1.43 | 1.50 | 4.9% |
| Total Emission (g) | | 3.63 | 3.71 | 2.2% |
| Total Emission/VMT (g/mile) | | 0.001296 | 0.001287 | -0.7% |

### 2. Network-wise Grid-based Concentration

To present the difference in traffic-related concentration, we firstly compare the estimated concentration profiles of both NOx and $PM_{2.5}$ (over the time) under two pickup scenarios at the location of the bus station. Then, we show the concentration profiles at 1.2m estimated by the grid-based dispersion model. Figure 9 displays the concentration levels at the central bus station. The figure shows that concentrations at the bus station are generally higher with the centralized pickup strategy compared to the distributed scenario. Specifically, mean concentrations of NOx and $PM_{2.5}$ throughout the centralized pickup strategy were 17% and 64% higher than those observed with the distributed pickup strategy, respectively.

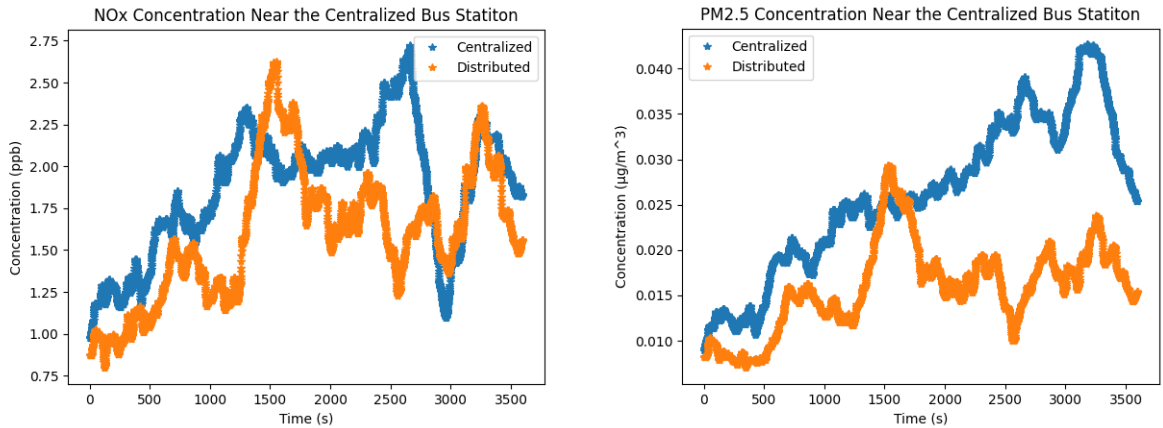

(a) NOx concentration profile  (b) $PM_{2.5}$ concentration profile

**Figure 9. Comparison of pollutant concentration profiles between two pickup scenarios at the central station.**

Furthermore, we show a typical screenshot of the network-wise concentration difference of both NOx and $PM_{2.5}$ between two pickup strategies (i.e., $conc_{central} - conc_{distributed}$) at the same time step in Figure 10. As can be observed from the figure, the largest difference in concentration between these two strategies occurs at or around the location of the central station.

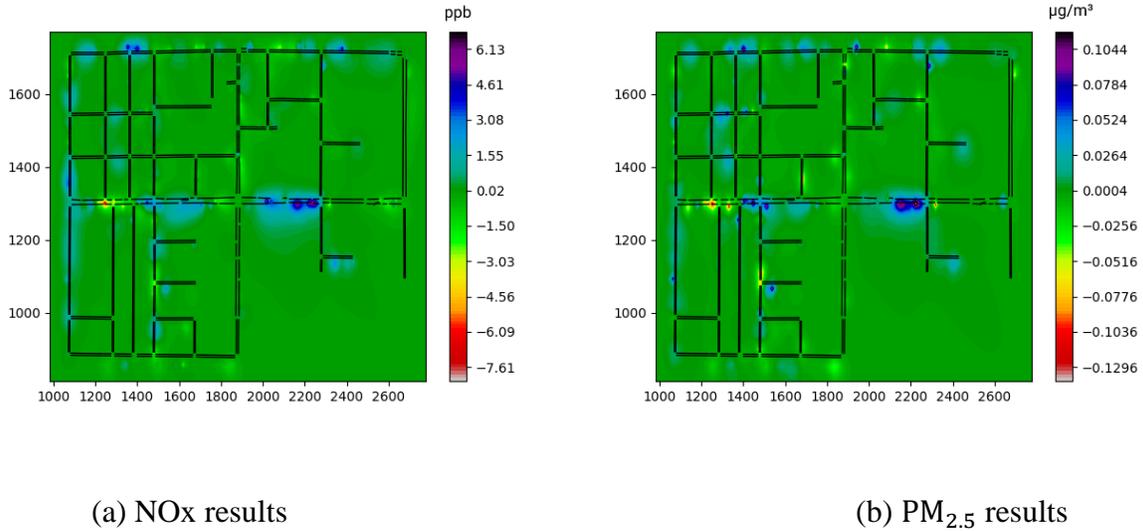

(a) NOx results　　　　　　　　　　　　　　(b) PM$_{2.5}$ results

**Figure 10. The difference of network-wise pollutant concentration (at 1.2m) between the centralized scenario and the distributed scenario.**

3. **Pedestrian Exposure**

Figure 11 displays the cumulative inhale mass (AIM) of PM$_{2.5}$ for a typical pedestrian, plotted against the dimensions of the map and time. The color-coded scatter plot indicates the position and AIM value of the pedestrian at each point in time, allowing for a better visualization of the pedestrian's trajectory and the associated PM$_{2.5}$ exposure over time. Once the pedestrian was picked up by a taxi, the color of the scatter plot stopped changing, indicating that the inhale mass was no longer accumulating as the pedestrian was no longer exposed to outdoor PM$_{2.5}$ while in the taxi.

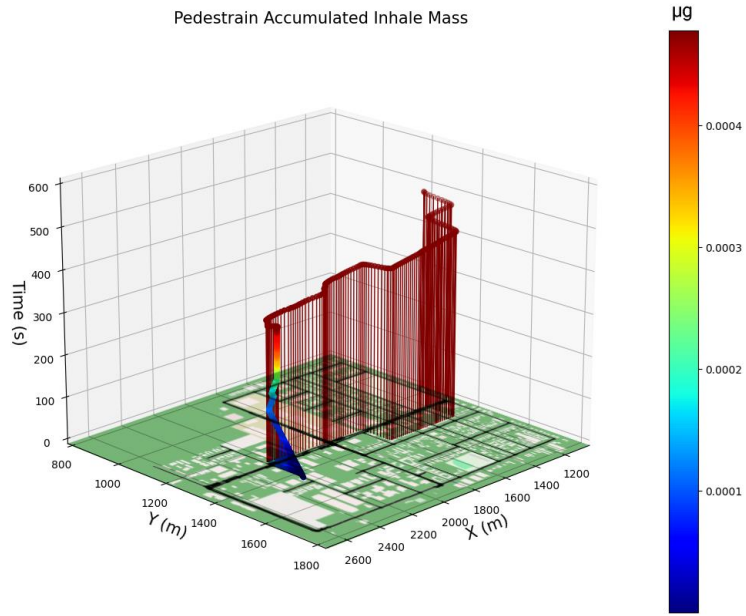

**Figure 11. Pedestrian AIM of $PM_{2.5}$ over time**

Table 3 summarizes the key statistics (e.g., maximum, minimum, median and mean) of the distributions of pedestrians' AIM of $PM_{2.5}$. The maximum AIM when applying the distributed pickup strategy is 33.2% lower than that of the centralized strategy. The median AIM and mean AIM in the distributed scenarios are close and lower than the values of the centralized scenario. The cancer risk is calculated based on Equation (6-8) and shown in Table 3.

**Table 3. Comparison of Human Exposure to $PM_{2.5}$ with Different Pickup Strategies**

| Scenario | Centralized | Distributed | Difference |
| --- | --- | --- | --- |
| Total Number of Pedestrians | 240 | 240 | 0% |
| Maximum Inhale Mass (μg) | 0.0066 | 0.0045 | -33.2% |
| Minimum Inhale Mass (μg) | 0 | 0 | 0% |
| Median Inhale Mass (μg) | 0.0028 | 0.0017 | -39.5% |
| Average Inhale Mass (μg) | 0.0027 | 0.0018 | -33.4% |
| Cancer Risk (in one million) | 0.016 | 0.011 | -33.4% |

## CONCLUSIONS AND FUTURE WORK

This study proposes an integrated analysis, modeling and simulation (AMS) platform for estimating traffic-related health impacts in a microscopic manner. Besides the traffic model and emission models, we introduce a 3D grid-based concentration model and a human exposure model as the key components of this platform. A case study on the evaluation of different curbside management strategies has shown the effectiveness of the AMS platform. A 5x5 grid size was chosen as it provided a good balance between spatial resolution and computational efficiency for obtaining reliable simulation results. It turns out that the centralized pickup strategy has more adverse effects on health in terms of human exposure to $PM_{2.5}$, compared to the distributed pickup strategy. The difference in cancer risk, on average, can be reduced by as much as 33.4%. The reasons may include a) relatively higher concentration due to the bottleneck created by the for-service vehicles; and b) longer waiting time for pedestrians in the centralized pickup scenario.

For future directions, we will apply this tool to investigate the health impacts of different roadway design and Transportation Systems Management and Operations (TSMO) strategies. In addition, we will conduct more field experiments to collect data for dispersion model calibration and keep improving the model accuracy by considering more realistic aerodynamic effects (e.g., turbulence) in the field of transportation.

## ACKNOWLEDGMENT

This research was funded by the Center for Advancing Research in Transportation Emissions, Energy, and Health (CARTEEH), project number 05-49-UCR. The authors would like to acknowledge the California Air Resource Board for providing PEAQS to measure CO2 concentration. The contents of this paper reflect only the views of the authors, who are responsible for the facts and the accuracy of the data presented herein.